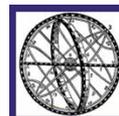

# Sundial-water clock of the Bronze Age (Northern Black Sea Region)

**Vodolazhskaya L.N. [1,*], Novichikhin A.M. [2], Nevsky M.Yu. [3]**

[1] Southern Federal University (SFU), str. Zorge, 5, Rostov-on-Don, 344090, Russian Federation;
E-mail: larisavodol@gmail.com

[2] Anapa Archaeological Museum, Naberezhnaya st., 4, Anapa, 353410, Russian Federation; E-Mail: yazamat10@yandex.ru

[2] Southern Federal University (SFU), str. Zorge, 5, Rostov-on-Don, 344090, Russian Federation;
E-mail: mnevsk@gmail.com

**Abstract**

This article presents the results of a study of signs on a Bronze Age slab discovered in the vicinity of a heavily plowed mound near the settlement of Pyatikhatki. The slab belongs to the Dolmen archaeological culture. In the course of this research, it was proved that the Pyatikhatki slab is a unique measuring tool, that combines elements of sundial and water clock. It has all the cup marks of an analemmatic sundial, except for the presence of a precise analemma, which at that time, it seems, could not yet be built. The vertical gnomon moved only along the north-south line, and the time was determined approximately. Most likely, this was due to the lack of accuracy in measuring time using the water clock of that era. It is possible that it was their imperfection that was the incentive for the development of a new type of watch - a sundial, which would allow measuring time over a long period with a higher accuracy.

In the autumn of 1982, a slab with cup marks and grooves was found near the settlement Pyatikhatki of Anapsky district, in the Northern Black Sea region (Fig. 1) (Novichikhin, 1995). It was removed from a heavily plowed mound. The finding was handed over to the Anapa Archaeological Museum, where it is exhibited[1].

The finding is a trapezoidal weathered limestone slab. On one of the flat sides, cup marks are knocked out, with a diameter of 2.5 to 11.5 cm and a depth of 1 to 4.6 cm, many of which are connected to each other by shallow grooves with a depth of 1 to 3 cm. The edges and inner surface of large cup marks (11.5-5.5 cm in diameter and 4.6-3 cm deep) are smoothed. Part of the slab is broken off. After being studied, the slab was attributed to the Dolmen archaeological culture of the Bronze Age and dated in the range from 2500 to 1500 BC.

The main feature of the slab is the cup marks arranged in a semicircle. To this day in two neighboring regions - Rostov and Donetsk - three slabs of fine-grained sandstone with cup marks located in a circle have been discovered, which belongs to the Srubnaya culture and dates back to the Late Bronze Age (approximately, XV-XII centuries BC), and also a slab of coarse-grained sandstone with a semicircular groove, which is approximately dated and attributed to the Yamnaya culture.

---

[1] The inventory number of the museum storage KM 7265/5

A slab with cup marks - from the Tavria-1 burial ground of the Rostov region is kept on the territory of the Archaeological Museum-Reserve "Tanais" (Rostov region, Russia) (Larenok, 1998, p. 62) and two slabs with cup marks - from the Rusin Yar and Popov Yar-2 mound groups, Donetsk region (Polidovych, Usachuk, 2013; Polidovich, Usachuk, 2015, pp. 444, 455; Polidovich, Usachuk, Kravchenko, Podobed, 2013, pp. 36–135), are located on the territory *of the Konstantinovskiy city museum* (Donetsk region, Ukraine) and in the Donetsk Region History Museum (Donetsk Region, Ukraine), respectively.

Slab with a groove having a shape of a semicircle, was found near the burial mound 1 Varvarinsky I Rostov region (Feifert, 2015, pp. 27-28) and is stored in the Azov historical and archaeological and paleontological museum-reserve (Rostov region., Russia).

Studies of slabs showed that cup marks and groove on these slabs constitute a hour markers and ellipse of "dial" analemmatic sundial respectively (Vodolazhskaya LN, 2013; Vodolazhskaya, Larenok, Nevsky, 2014; Vodolazhskaya, Larenok, Nevsky, 2016a; Vodolazhskaya, Larenok, Nevsky, 2016b).

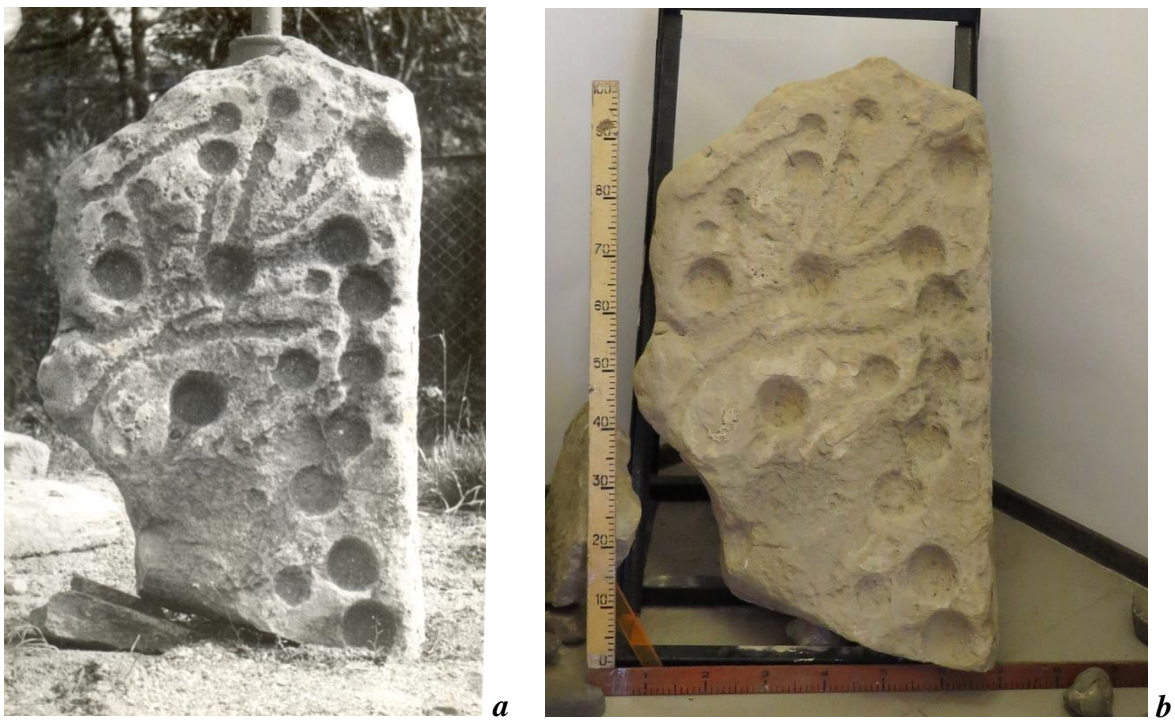

**Figure 1.** Slab from the settlement Pyatikhatki: a - in the courtyard of the Anapa Archaeological Museum. Photo from 1984; b - in the modern exposition of the Anapa Archaeological Museum.

*In order* to analyze the location of cup marks and grooves on the surface of the slab from Pyatikhatki, a photograph was taken of a vertically installed slab with horizontally and vertically mounted photometers located as close as possible to the slab surface under study (Fig. 2a).

Then, the distortions that appeared at the edges of the photo were corrected with the help of the "Transform" tool of the computer program Photoshop CS5 (fig. 2b). On the basis of the corrected photograph, the surface of the slab was drawn in the form of a plan-diagram, which most accurately reflects the location of the cup marks and grooves (Fig. 3).

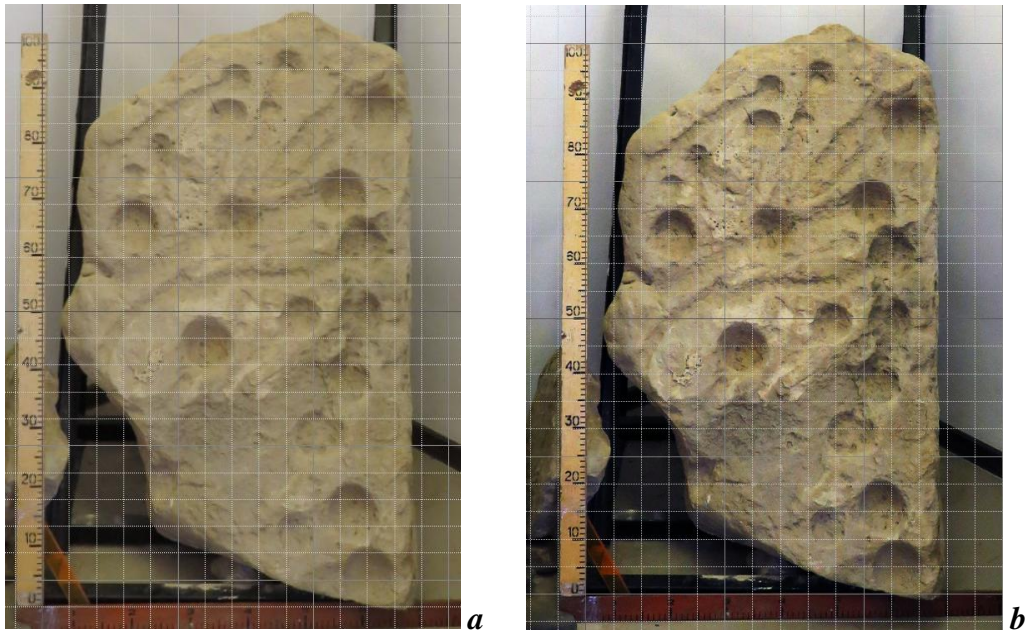

**Figure 2.** Photo of a slab from settlement Pyatikhatki: a - without correction; b - corrected photograph (removed distortions at the edges of the photograph, increased contrast).

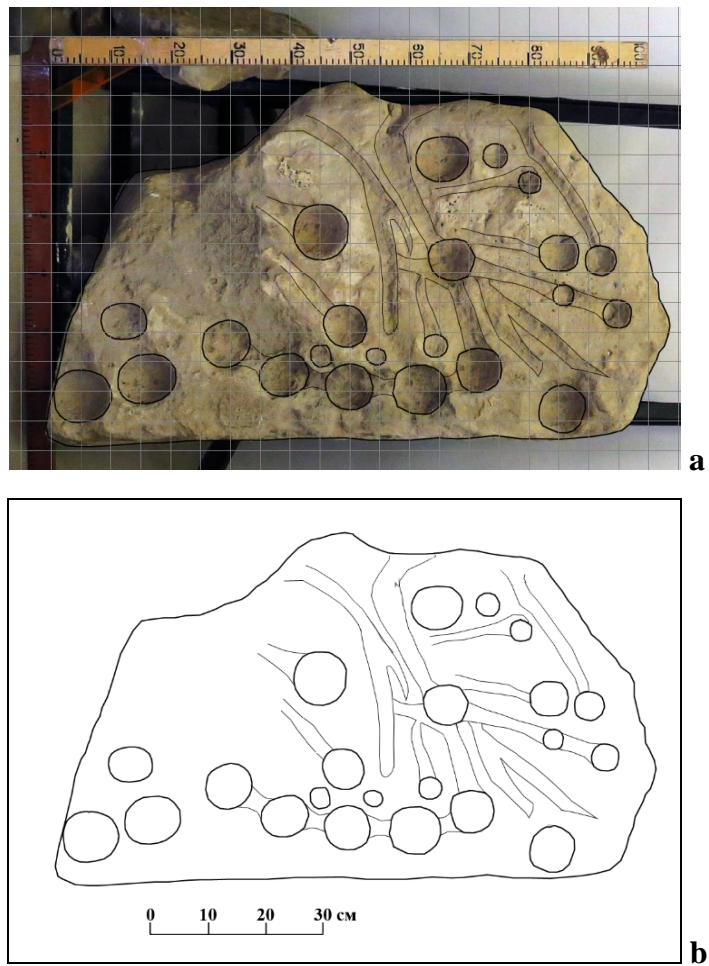

**Figure 3.** Plan-scheme of the surface of the slab from the settlement Pyatikhatki: a - photograph of the slab with a schematic drawing; b - schematic drawing of the slab surface.

The cup marks on the slab from Pyatikhatki are located quite chaotically and only a small fragment of an ellipse of five cup marks is clearly distinguishable, so it was not possible to immediately determine and measure the semiaxes of the ellipse of the supposed "dial" of the analemmatic clock. Initially, to solve this problem an ellipse of the maximum possible size was considered, with the value of the semi-minor axis *m* approximately equal to half the width of the slab m≈25 cm.

The slab belongs to the Dolmen culture and dates from 2500 BC to 1500 BC, therefore the calculations of the hour markers of the analemmatic sundial were originally made by us for 1500 BC for the geographical coordinates of the place of detection 44°58′ N, 37°18′ E (settlement Pyatikhatki, Krasnodar Krai).

The calculations were carried out according to formulas 1-6 (Savoie, 2009, p. 121):

$$M = \frac{m}{\sin \varphi}, \tag{1}$$

$$x = M \cdot \sin H, \tag{2}$$

$$y = M \cdot \sin \varphi \cdot \cos H, \tag{3}$$

$$Z_{ws} = M \cdot tg\delta_{ws} \cdot \cos \varphi, \tag{4}$$

$$Z_{ss} = M \cdot tg\delta_{ss} \cdot \cos \varphi, \tag{5}$$

$$H' = arctg(tgH/\sin\varphi), \text{ at } t \in [6; 18] \tag{6}$$

$$H' = arctg(tgH/\sin\varphi) - 180°, \text{ at } t \in [0; 6[$$

$$H' = arctg(tgH/\sin\varphi) + 180°, \text{ at } t \in ]18; 24],$$

$$\text{where } H = 15° \cdot (t - 12),$$

where *x* - the coordinate of a point along the *X* axis for an analemmatic sundial, *y* - the coordinate of a point along the *Y* axis for an analemmatic sundial, *m* - the semi-minor axis of the ellipse, *M* - the semi-major axis of the ellipse, *φ* - the latitude of the area, *t* - the true local solar time, *H* - the hour angle of the Sun, *H'* - the angle between the noon line and the hour line on the clock relative to the center of coordinates (center of the ellipse), $\delta_{ws}=-\varepsilon$ - declination of the Sun on the day of the winter solstice, $\delta_{ss}=\varepsilon$ - declination of the Sun on the day of the summer solstice, $y=Z_{ws}$ - on the day of the winter solstice, $y=Z_{ss}$ - on the day of the summer solstice (Fig. 4). On the days of the equinox $\delta_{eq}=0$.

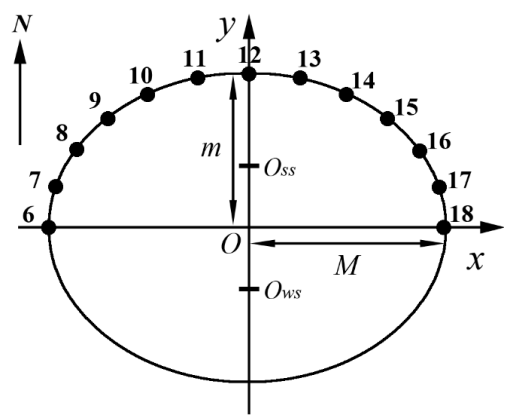

**Figure 4.** Coordinate plane with hour markers from 6 to 18 o'clock. M is the semi-major axis of the ellipse, *m* is the semi-minor axis of the ellipse, *O* is the center of the ellipse, $O_{ws}$ is the position of the gnomon on the winter solstice for analemmatic clock, $O_{ss}$ is the position of the gnomon on the summer solstice for the analemmatic clock.

Using the calculated coordinates, hour markers were plotted on a coordinate plane projected onto the drawing of the slab surface. Then the location of the center of the ellipse was selected, allowing approximately to combine the large cup marks located in a semicircle on the slab and the arc of the ellipse formed by the clock marks. For the most accurate match, it was necessary to shift the center of the ellipse to the southern edge of the slab and increase the semi-minor axis *m* to 29 cm. Then, again, using formulas 1-6, the coordinates of hour markers were calculated, but already for m≈29 cm. In this case, the bridges between the large cup marks coincided as closely as possible with the location of the clock marks, and not the large cup marks themselves, oddly enough (Fig. 5).

Calculated for the semi-minor axis *m = 29 cm* the semi-major axis is *M≈41 cm*, the magnitude of the displacement of the gnomon along the Y-axis on the winter solstice is $Z_{ws}$≈*-12.9 cm*, and on the summer solstice $Z_{ss}$≈*12.9 cm*. That is, to correctly measure the time, the gnomon had to be shifted on the day of the summer solstice by ≈*12.9 cm* north of the center of coordinates (point O), and on the day of the winter solstice by ≈*12.9 cm* to the south of the center of coordinates (Fig. 5).

The results of calculating the *x* and *y* coordinates of the hour markers of the analemmatic sundial for the latitude of the Pyatikhatki are given in Table 1.

**Table 1.** Calculated coordinates of hour markers of the analemmatic sundial for latitude 44°58′ N: *H* - the hour angle of the Sun, *H'* - the angle between the noon line and the hour line on the sundial, *t* - the mean solar time, *x* - the coordinate of the hour markers along the X axis, *y* - the y-coordinate of the clock marks.

|  | *t* | | | | | | | | | | | | |
|---|---|---|---|---|---|---|---|---|---|---|---|---|---|
|  | 0 | 1 | 2 | 3 | 4 | 5 | 6 | 7 | 8 | 9 | 10 | 11 | 12 |
| $H, (^0)$ | -180,0 | -165,0 | -150,0 | -135,0 | -120,0 | -105,0 | -90,0 | -75,0 | -60,0 | -45,0 | -30,0 | -15,0 | 0,0 |
| $H', (^0)$ | -180,0 | -159,2 | -140,8 | -125,2 | -112,2 | -100,7 | -90,0 | -79,3 | -67,8 | -54,8 | -39,2 | -20,8 | 0,0 |
| *x*, (cm) | 0,0 | -10,6 | -20,5 | -29,0 | -35,5 | -39,6 | -41,0 | -39,6 | -35,5 | -29,0 | -20,5 | -10,6 | 0,0 |
| *y*, (cm) | -29,0 | -28,0 | -25,1 | -20,5 | -14,5 | -7,5 | 0,0 | 7,5 | 14,5 | 20,5 | 25,1 | 28,0 | 29,0 |

|  | *t* | | | | | | | | | | | | |
|---|---|---|---|---|---|---|---|---|---|---|---|---|---|
|  | 12 | 13 | 14 | 15 | 16 | 17 | 18 | 19 | 20 | 21 | 22 | 23 | 24 |
| $H, (^0)$ | 0,0 | 15,0 | 30,0 | 45,0 | 60,0 | 75,0 | 90,0 | 105,0 | 120,0 | 135,0 | 150,0 | 165,0 | 180,0 |
| $H', (^0)$ | 0,0 | 20,8 | 39,2 | 54,8 | 67,8 | 79,3 | 90,0 | 100,7 | 112,2 | 125,2 | 140,8 | 159,2 | 180,0 |
| *x*, (cm) | 0,0 | 10,6 | 20,5 | 29,0 | 35,5 | 39,6 | 41,0 | 39,6 | 35,5 | 29,0 | 20,5 | 10,6 | 0,0 |
| *y*, (cm) | 29,0 | 28,0 | 25,1 | 20,5 | 14,5 | 7,5 | 0,0 | -7,5 | -14,5 | -20,5 | -25,1 | -28,0 | -29,0 |

Since the results obtained made it possible to explain the location of the cup marks and grooves on the slab only partially, it was decided to additionally calculate the analemma for the gnomon - the coordinates of the exact location of the gnomon throughout the year in order to test the hypothesis about the possible coincidence of some grooves with the shadow from the gnomon on astronomically significant days of the year.

To accurately measure the time in the analemmatic sundial, the gnomon must be moved along the analemma throughout the year. The analemma has the form of a figure of eight and reflects the position of the Sun at the same mean solar time of the day (in the case of analemmatic sundial, at noon) throughout the year. The change in the position of the Sun during the year occurs due to the discrepancy between the true solar time and the mean solar time. The analemma could have been constructed in ancient times empirically while measuring time with the help of a sundial and a water clock.

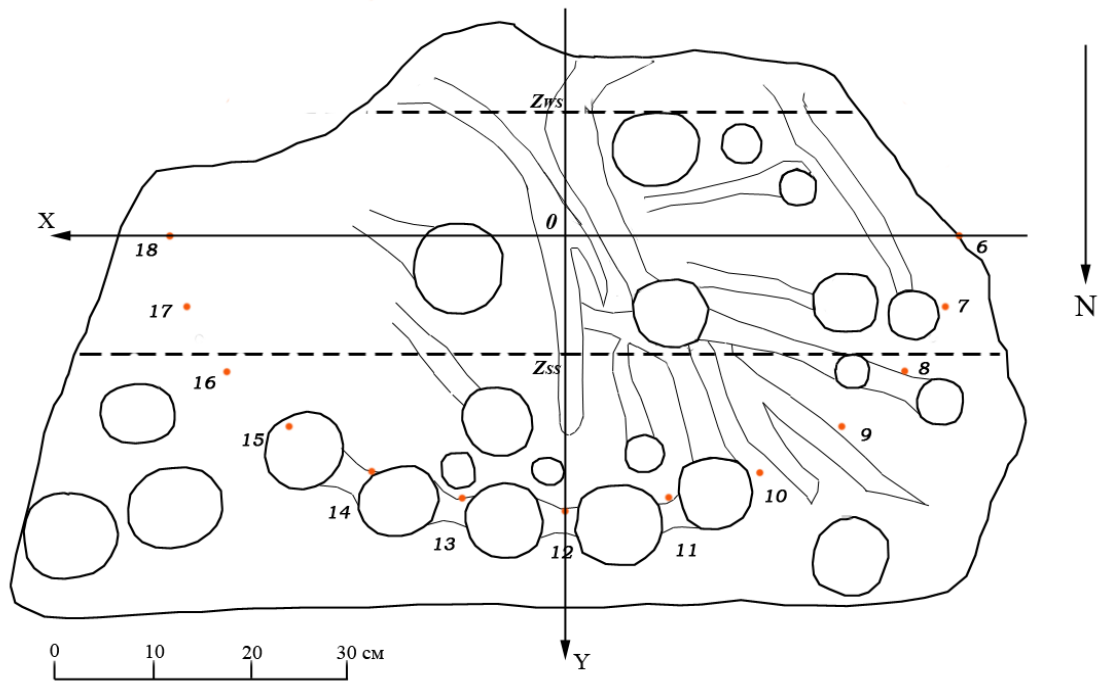

**Figure 5.** Hour markers of the analemmatic sundial plotted on the slab surface plan. Hour markers are highlighted in orange and signed with the number of the corresponding hour (from 6 to 18 hours). *N* - direction to the True North.

The sundial shows true solar time. Mean solar time can be measured with a water clock, which was most likely used during the Bronze Age to mark analemmatic sundial by setting the length of the hour. In order to achieve the coincidence of the readings of the analemmatic sundial with the readings of the water clock, it was necessary for the gnomon to move in accordance with the solar analemma during the year.

The unevenness of the daily motion of the Sun is due to the ellipticity of the Earth's orbit around the Sun and the inclination of the Earth's axis to the plane of the ecliptic. The duration of an average solar day is the average value of the duration of a true solar day per year. The difference between true solar time and mean solar time at the same moment is called the equation of time[2] and is calculated using formula 7:

$$\eta = T_s - T_m ,  \qquad (7)$$

where $\eta$ - the equation of time (minutes), $T_s$ - true solar time, $T_m$ - mean solar time. If the value of the equation of time is negative, then the true solar time lags behind the mean solar time, determined, for example, by the water clock, and if it is positive, then it is ahead.

Calculations of the equation of time were carried out by us using the astronomical program HORIZONS System[3]. The calculation results for each month, starting from the day of the vernal equinox, are presented in Table 2.

---

[2] Sometimes an inverted time equation is used, which is equal to the difference between mean time and true solar time
[3] http://ssd.jpl.nasa.gov/?horizons

The movement of the gnomon in the analemmatic sundial along the *Y* axis was calculated using formula 8, and along the *X* axis it was approximately calculated using formula 9, which we obtained:

$$Z_y = M \cdot tg\delta \cdot \cos\varphi, \tag{8}$$

$$Z_x = \eta \cdot \frac{M \cdot \sin(15°)}{60}, \tag{9}$$

where $Z_y$ - the displacement of the gnomon along the *Y* axis, $Z_x$ - the displacement of the gnomon along the *X* axis, *M* - the semi-major axis of the ellipse, *δ* - the declination of the Sun, *φ* - the latitude of the area, *η* - the equation of time.

The equinox, solstice and declination dates were calculated by us using the astronomical program RedShift 7.

The calculation results are presented in Tables 2, 3, 4. The drawing of the analemma is shown in Figure 6.

**Table 2.** Coordinates of analemma points for 2500 BC and 1500 BC for latitude 44°58′ N: $Z_y$ - displacement of the gnomon along the *Y* axis, $Z_x$ - displacement of the gnomon along the *X* axis, *δ* - declination of the Sun, *η* - equation of time.

| Point number on the analemma | Date | 2500 BC | | | | 1500 BC | | | |
|---|---|---|---|---|---|---|---|---|---|
| | | *δ*, ° | *η*, min | $Z_x$, cm | $Z_y$, cm | *δ*, ° | *η*, min | $Z_x$, cm | $Z_y$, cm |
| 1 | April (spring equinox) /3 April 1500 BC 11 April 2500 BC ./ | 0,02 | -6 | -1,1 | 0,0 | -0,05 | -7 | -1,2 | 0,0 |
| 2 | May /3 May 1500 BC 11 May 2500 BC / | 11,22 | 7 | 1,2 | 5,8 | 11,13 | 5 | 0,9 | 5,7 |
| 3 | June /3 June 1500 BC 11 June 2500 BC / | 20,18 | 12 | 2,1 | 10,7 | 20,07 | 10 | 1,8 | 10,6 |
| 4 | July /3 July 1500 BC 11 July 2500 BC / | 23,93 | 7 | 1,2 | 12,9 | 23,83 | 6 | 1,1 | 12,8 |
| 5 | August /3 August 1500 BC 11 August 2500 BC / | 21,18 | -2 | -0,4 | 11,3 | 21,20 | -2 | -0,4 | 11,3 |
| 6 | September /3 September 1500 BC 11 September 2500 BC / | 12,42 | -4 | -0,7 | 6,4 | 12,67 | -2 | -0,4 | 6,5 |
| 7 | October /3 October 1500 BC 11 October 2500 BC / | 0,63 | 1 | 0,2 | 0,3 | 1,12 | 4 | 0,7 | 0,6 |
| 8 | November /3 November 1500 BC 11 November 2500 BC / | -11,75 | 6 | 1,1 | -6,0 | -11,22 | 9 | 1,6 | -5,8 |
| 9 | December /3 December 1500 BC 11 December 2500 BC / | -20,73 | 2 | 0,4 | -11,0 | -20,37 | 6 | 1,1 | -10,8 |

| | | | | | | | | |
|---|---|---|---|---|---|---|---|---|
| 10 | January /3 January 1500 BC 11 January 2500 BC / | -23,92 | -11 | -1,9 | -12,9 | -23,82 | -8 | -1,4 | -12,8 |
| 11 | February /3 February 1500 BC 11 February 2500 BC / | -19,90 | -20 | -3,5 | -10,5 | -19,98 | -19 | -3,4 | -10,6 |
| 12 | March /3 March 1500 BC 11 March 2500 BC / | -11,60 | -18 | -3,2 | -6,0 | -11,72 | -18 | -3,2 | -6,0 |

The equation of time in 1500 BC is equal to zero on April 19-20 and July 24-27 (point Z on the plan-diagram) (Fig. 6). On these days, the time shown by the sundial and water clock coincided and the movement of the gnomon along the X axis was not required.

**Table 3.** Coordinates of the analemma points on the days of the solstices and equinoxes in 1500 BC for geographical coordinates 44°58′ N, 37°18′ E: $\delta$ - declination of the Sun, $\eta$ - equation of time, $Z_x$ - displacement of the gnomon along the *X* axis, $Z_y$ - displacement of the gnomon along the *Y* axis.

| Designation points | Astronomical phenomenon | Date | $\delta$ | $\eta$, min | $Z_x$, cm | $Z_y$, cm |
|---|---|---|---|---|---|---|
| VE | Vernal equinox | 3 April | 0° | -7 | -1,2 | 0,0 |
| SS | Summer solstice | 6 July | 23,90° | 5 | 0,9 | 12,9 |
| AE | Aautumn equinox | 6 October | 0° | 5 | 0,9 | 0,0 |
| WS | Winter solstice | 2 January | -23,90° | -7 | -1,2 | -12,9 |
| | | | | | | |

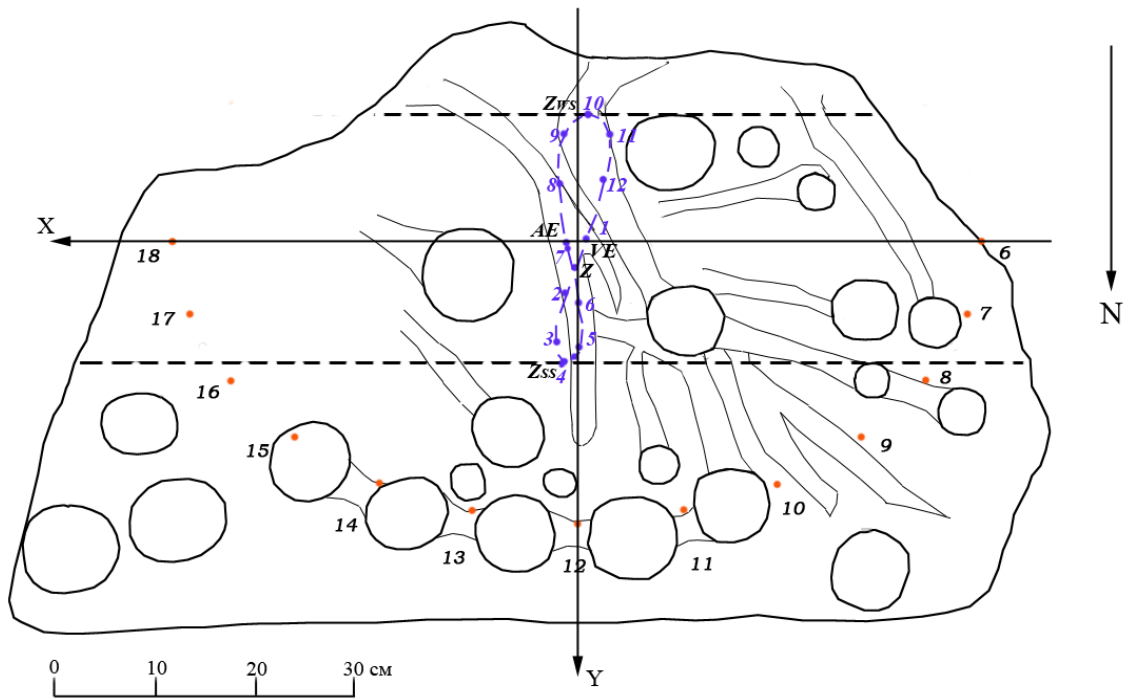

**Figure 6.** Analemma for 1500 BC, built in the coordinate plane on the plan-scheme of the slab from Pyatikhatki.

**Table 4.** Coordinates of the analemma points on the days of the solstices and equinoxes in 2500 BC for geographical coordinates 44°58′ N, 37°18′ E: $\delta$ - declination of the Sun, $\eta$ - equation of time, $Z_x$ - displacement of the gnomon along the *X* axis, $Z_y$ - displacement of the gnomon along the *Y* axis.

| Designation points | Astronomical phenomenon | Date | $\delta$ | $\eta$, min | $Z_x$, cm | $Z_y$, cm |
|---|---|---|---|---|---|---|
| VE | Vernal equinox | 11 April | 0° | -6 | -1,2 | 0,0 |
| SS | Summer solstice | 14 July | 23,98° | 6 | 0,9 | 12,9 |
| AE | Aautumn equinox | 13 October | 0° | 2 | 0,9 | 0,0 |
| WS | Winter solstice | 9 January | -23,98° | -10 | -1,2 | -12,9 |

Tables 2 - 4 show that the values of the equation of time and the coordinates of the analemma for 2500 BC and 1500 BC differ insignificantly, as well as the coordinates of clock marks, therefore, further analysis was carried out by us only for 1500 BC.

For a more accurate analysis of the possible connection of the grooves with the hour lines, we built hour lines for the times of sunrise and sunset on astronomically significant days 1500 BC. Sunrise and sunset times for these days were calculated using the RedShift 7 Advanced software. The calculation results are presented in Table 5 and Figure 7.

**Table 5.** Coordinates of hour markers of analemmatic sundial for geographical coordinates 44°58′ N, 37°18′ E for sunrise and sunset on astronomically significant days 1500 BC: t - mean solar time, x - coordinate of hour markers on the X-axis, y - the coordinate of the hour markers along the Y-axis.

| Analemma point | Date | Sunrise | | | Sunset | | |
|---|---|---|---|---|---|---|---|
| | | t | x, cm | y, cm | t | x, cm | y, cm |
| VE | Spring equinox 3 April | 5:58 | -41,0 | -0,3 | 18:03 | 41,0 | -0,4 |
| Z | $\eta=0$ in spring 19 April | 5:33 | -40,8 | -3,4 | 18:28 | 40,7 | -3,5 |
| SS | Summer solstice 6 July | 4:11 | -36,5 | -13,3 | 19:49 | 36,5 | -13,3 |
| Z | $\eta=0$ in summer 24 July | 4:17 | -37,0 | -12,6 | 19:43 | 37,0 | -12,6 |
| AE | Autumn equinox 13 October | 5:58 | -41,0 | -0,3 | 18:03 | 41,0 | -0,4 |
| WS | Winter solstice 9 January | 7:41 | -37,1 | 12,4 | 16:19 | 37,1 | 12,4 |

Figure 8 shows that each hour marker can be associated with one of the cup marks located along the perimeter of the slab. In the figure, these cup marks are designated L4 - L18. The numbers of these cup marks correspond to the numbers of the hour markers (the hour corresponding to the marker). Most of the grooves extending from these cup marks (or located close to them) are aligned with the corresponding hour lines and in some cases coincide quite well with them. This applies to the grooves coming from the cup marks: L7, L8, L9 (presumably located on the slab chip), L10, L11 (with some deviation), L12, L13. This fact suggests that the marking of the slab was carried out either on the day of the vernal equinox, or, first of all, for the day of the vernal equinox.

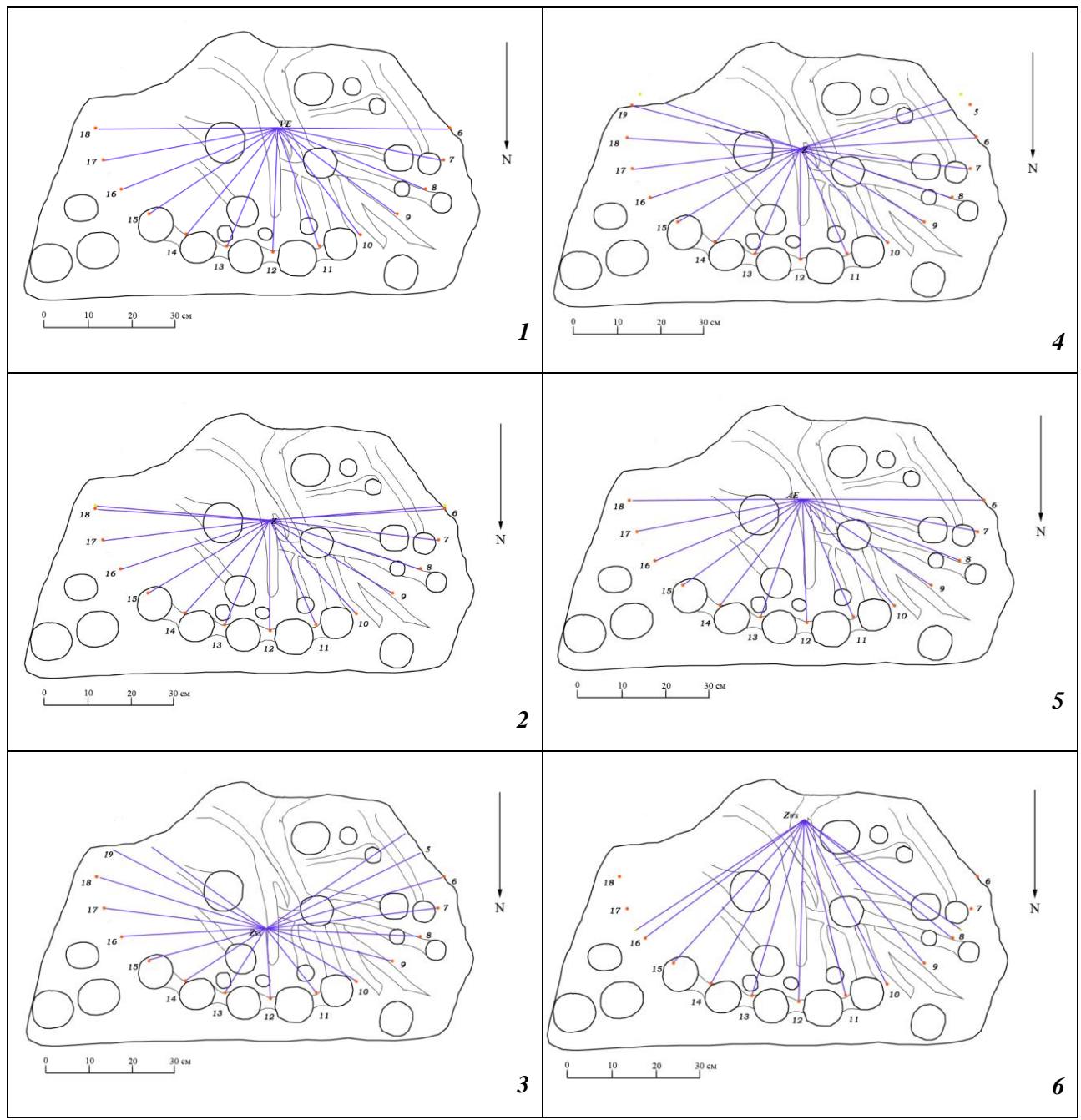

**Figure 7.** Plan-scheme of a slab with hour markers and hour lines for astronomically significant days of 1500 BC: 1 - the day of the vernal equinox; 2 - day when η = 0 (April 19); 3 - the day of the summer solstice; 4 - day when η = 0 (July 24); 5 - the day of the autumnal equinox; 6 - the day of the winter solstice.

The most interesting are the hour lines for the vernal equinox (Fig. 8).

The groove from cup mark L5 also almost coincides with the five o'clock hour line, but on the vernal equinox the sun rises at six o'clock in the morning. This may indicate that the marking of the slab was carried out according to a previously prepared template, on which earlier hour markers were also marked. Cup marks L11 - L15 correspond most closely to the ellipse of hour markers of the analemmatic sundial. Therefore, we believe that the purpose of creating the slab of Pyatikhatki was to measure time precisely at these hours. The marking of the slab and the drawing of cup marks and grooves on it also began from these cup marks. Other cup marks were already applied approximately, without strict reference to hour markers. Cup marks L16 - L18 correspond to the remaining hour markers only in number, and are not located next to them, but only in the corresponding corner of the slab. A similar situation is observed with cup mark L10.

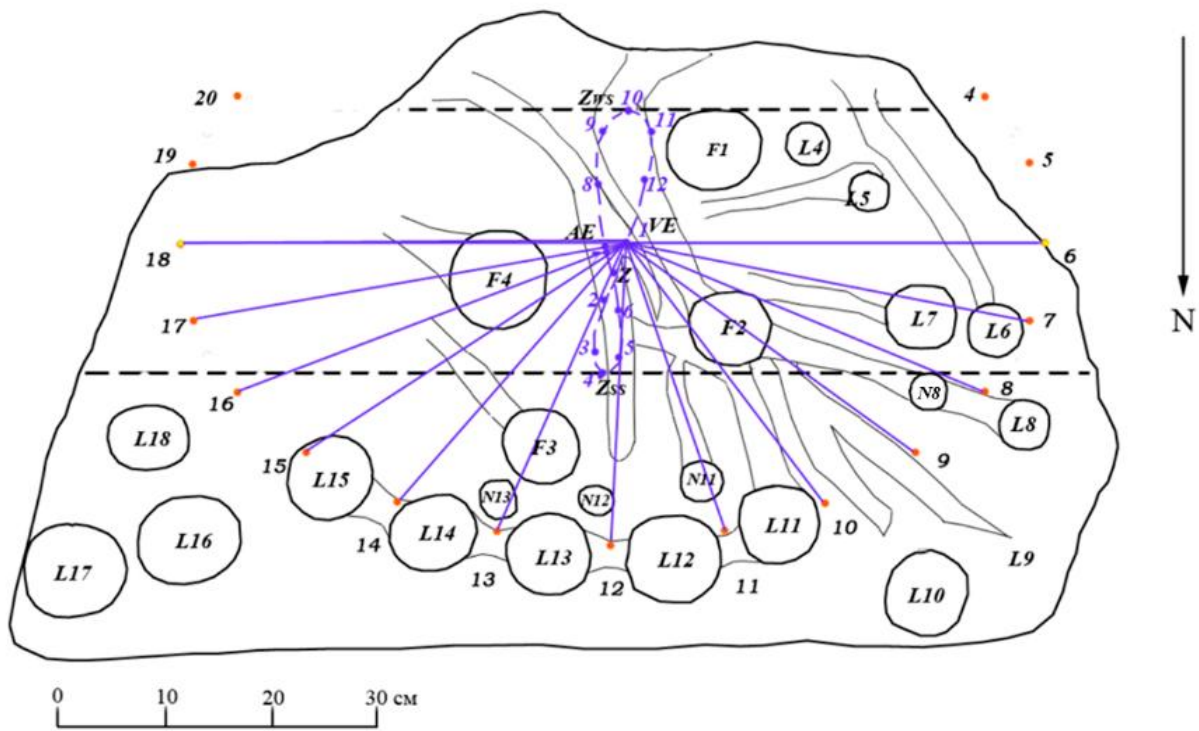

**Figure 8.** Plan-scheme of a slab with hour markers and lines for the vernal equinox.

To be able to position the cup marks L4 - L8 closer to the hour markers, they had to be made smaller. The hour markers at 4-6 o'clock are already outside the slab or directly on its edge, so the cup marks L4 - L6 were placed closer to the center of the slab, and the cup mark corresponding to 7 o'clock was "given" to 6 o'clock (noting this with a groove going towards the six o'clock line) , and for 7 hours, cup mark L7 was plotted closer to the center.

Large cup marks F1 - F4 are located in the center of the slab, surrounding the area within which the gnomon was supposed to move, therefore, most likely, they were associated with the gnomon mount. Apparently cup marks F1, F2 and F4 were used to mount the gnomon in the spring and autumn, and in the summer - cup marks F2 - F4.

In addition, the southern edge of the F1 cup mark and the northern edge of the F2 cup mark coincide with the $Z_{ws}$ and $Z_{ss}$ lines, which set the magnitude of the gnomon shift on the days of the winter and summer solstices, respectively. That is, we can say that the cup marks F1 and F2 mark these distances on the slab. The southern edge of cup mark F4 roughly corresponds to the X-axis (6 and 18 o'clock hour lines) and marks the position of the gnomon on the equinox days.

Calculations of the volume of cup marks L11 - L15 using the formula for the volume of a spherical segment showed that their volumes are close to the volume of water for measuring the time of one hour using the water clock of the Bronze Age, found in the Donetsk region, equal to $136.2 \pm 21.7$ cm$^3$ (tabl. 6 ) (Vodolazhskaya, Usachuk, Nevsky, 2015).

**Table 6.** Volume of cup marks L11 - L15

| Cup mark | Diameter, cm | Depth, cm | Volume, cm$^3$ |
|---|---|---|---|
| L11 | 9,5 | 4,5 | 207,1 |
| L12 | 9,5 | 4,2 | 187,6 |
| L13 | 8,0 | 4,0 | 134,0 |
| L14 | 8,0 | 3,8 | 124,2 |
| L15 | 9,0 | 4,0 | 160,7 |

That is, when marking the slab from Pyatikhatki, a similar water clock was used, and the cup marks L11 - L15 could themselves serve as a water clock for measuring time with a duration of one hour. However, the presence of grooves between these cup marks reduces their working volume, so they themselves were most likely not used to measure time as a water clock. They could accommodate vessels similar in shape and volume, or one and the same vessel, which was installed in turn, used as a water clock to measure one - the current hour.

The length of the groove between the central cup marks L12 and L13 is approximately equal to the distance on the X-axis between the vernal equinox VE (or the winter solstice point $Z_{ws}$) and the easternmost point 3 (or point 8) of the analemma (Fig. 9a). In fact, the length of the groove corresponds to the area in which the midday shadow of the gnomon would appear throughout the year, except for the two months following the winter solstice, if the gnomon were not moved along the X-axis (simplified diagram of an analemmatic sundial).

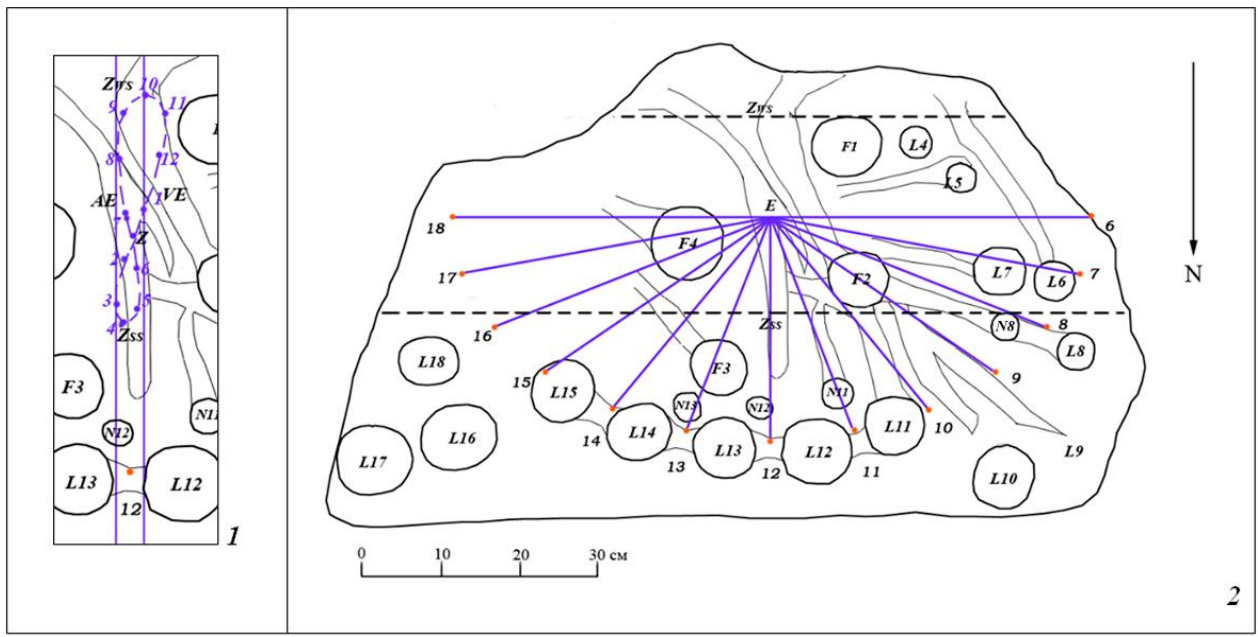

**Figure 9.** Simplified scheme of analemmatic sundial: *1* - projection of point VE and point 3 of the analemma on the region of the groove between cup marks L12 and L13; *2* – plan-scheme of a slab with hour lines for the days of the equinox with a gnomon set at the center of coordinates.

The projections of the westernmost points of the analemma (11 and 12) extend beyond the groove between cup marks L11 and L12. This may indicate that in the winter months following the winter solstice, when the number of sunny days is minimal, no time was measured with a sundial. In the case of a simplified scheme of analemmatic sundials, the gnomon had to be installed on the days of the equinox in the center of coordinates (corresponds to the bifurcation of the central groove) (fig.9b) and move throughout the year only along the Y axis. That is, from the day of the vernal equinox to the middle of December (the last days of November in our time) the gnomon was supposed to move along a groove that runs approximately from the bridge between cup marks L12 and L13 to the center of the slab and sharply turns in a southeast direction where it begins to narrow outside the analemma. During the winter months, the gnomon had to be rearranged even further north - to the adjacent groove, next to the F1 cup mark. The gnomon was not moved along it in the winter months, because after the winter solstice, the measurement of time, as noted above, most likely, was not performed due to bad weather conditions.

The lengths of the grooves between adjacent cup marks in the range L11 - L15 and the width of most of the grooves on the slab is about 3 cm. Probably, with the help of wide grooves, they tried to take into account the uneven motion of the Sun when measuring time by sundials, and to achieve at least an approximate coincidence of the results of time measurements by solar and water clock throughout the year. On the slab, next to cup marks L8, L11, L12 and L13, there are small cup marks: N8, N11, N12 and N13. Their diameter is about 3 cm, like the grooves. It seems that they were the first to reflect the idea of marking with a cup mark not the gap between the hour lines, but the hour mark itself. It was she who later formed the basis for the principle of marking the "dial" of the analemmatic sundial of the Bronze Age of the Northern Black Sea region with the help of small-diameter cup marks that corresponded to hour markers, and not to the intervals between them. So the marking of analemmatic sundial is located in the working range in accordance with hour markers with a diameter of about 3 ÷ 4 cm on a slab from the Tavria-1 burial ground (Vodolazhskaya, Larenok, Nevsky, 2014) and on a slab from the Popov Yar-2 mound group ( Polidovych, Usachuk, 2013; Vodolazhskaya, 2013; Vodolazhskaya, Larenok, Nevsky, 2016a).

Thus, the stove from Pyatikhatki is a unique measuring instrument that combines elements of sundial and water clock. It has all the cup marks of an analemmatic sundial, except for the presence of a precise analemma, which at that time, it seems, could not yet be built. The vertical gnomon moved only along the north-south line, and the time was determined approximately. Most likely, this was due to the lack of accuracy in measuring time using the water clock of that era. It is possible that it was their imperfection that was the incentive for the development of a new type of watch - a sundial, which would allow measuring time over a long period with a higher accuracy.